\documentclass[journal]{IEEEtran}

\usepackage{cite,citesort}
\usepackage{graphicx}
\usepackage{amsmath}
\usepackage{amssymb}
\usepackage{amsfonts}
\usepackage{algorithm}
\usepackage{algorithmic}
\usepackage{balance}
\usepackage{stfloats}
\usepackage{epstopdf}
\usepackage{subfigure}

\DeclareMathOperator*{\argmin}{argmin}

\makeatletter
\def\ScaleIfNeeded{%
\ifdim\Gin@nat@width>\linewidth \linewidth \else \Gin@nat@width
\fi } \makeatother

\newtheorem{theorem}{Theorem}

\newtheorem{proposition}{Proposition}

\title{Location-Based Beamforming and Physical Layer Security in Rician Wiretap Channels}
\begin{document}

\author{Chenxi~Liu,~\IEEEmembership{Student Member,~IEEE,} and Robert~Malaney,~\IEEEmembership{Member,~IEEE}
\thanks{The work of R. Malaney was supported by the Australian Research Council Discovery Project (DP120102607).}
\thanks{C. Liu and R. Malaney are with the School of Electrical Engineering and Telecommunications, The University of New South Wales, Sydney, NSW 2052, Australia (email:
chenxi.liu@student.unsw.edu.au; r.malaney@unsw.edu.au).}}


\maketitle

\begin{abstract}
We propose a new location-based beamforming (LBB) scheme for wiretap channels, where a multi-antenna source communicates with a single-antenna legitimate receiver in the presence of a multi-antenna eavesdropper. We assume that all channels are in a Rician fading environment, the channel state information from the legitimate receiver is perfectly known at the source, and that the only information on the eavesdropper available at the source is her location. We first describe how the optimal beamforming vector that minimizes the secrecy outage probability of the system is obtained, illustrating its dependence on the eavesdropper's location. We then derive an easy-to-compute expression for the secrecy outage probability when our proposed LBB scheme is adopted. We also consider the positive impact a friendly jammer can have on our beamforming solution, showing how the path to optimality remains the same. Finally, we investigate the impact of location
uncertainty on the secrecy outage probability, showing how our
solution can still allow for secrecy even when the source
only has a noisy estimate of  the eavesdropper's location. Our work demonstrates how a multi-antenna array, operating in the most general channel conditions and most likely system set-up, can be configured rapidly in the field so as to deliver an optimal physical layer security solution.
\end{abstract}

\section{Introduction}
\PARstart{P}{hysical} layer security has attracted significant research attention recently. Compared to the traditional upper-layer cryptographic techniques using secret keys, physical layer security safeguards wireless communications by directly exploiting the randomness offered by wireless channels without using secret keys, and thus has been recognized as an alternative for cryptographic techniques \cite{Yang_Mag}. The principle of physical layer security was first studied in \cite{wyner} assuming single-input single-output systems. It was shown that secrecy can only exist when the wiretap channel between the source and the eavesdropper is a degraded version of the main channel between the source and the legitimate receiver. Subsequently, this result was generalized to the case where the main channel and the wiretap channel are independent \cite{csiszar}.

More recently, implementing multi-input multi-output (MIMO) techniques at the source/legitimate receiver has been shown to significantly improve the physical layer security of wiretap channels \cite{khisti,wornell,chenxi,chenxi2,nan4,Zhang13,nan5,nan6,nan,nan3,jiangyuan,yansha_tcom1,yansha_tcom2}. In terms of MIMO techniques, beamforming \cite{khisti,wornell,chenxi,chenxi2,nan4,jiangyuan}, artificial noise \cite{Zhang13,nan5,nan6,yansha_tcom1}, and transmit antenna selection \cite{nan,nan3,yansha_tcom2} are just a few techniques that can be utilized to boost the physical layer security of wiretap channels. In \cite{khisti,wornell,chenxi,chenxi2,nan4,Zhang13,nan5,nan6,nan,nan3,jiangyuan,yansha_tcom1,yansha_tcom2}, it is assumed that the channel state information (CSI) from the eavesdropper is perfectly or statistically known at the source. This assumption, however, is unlikely to be valid in practice - especially when the eavesdropper is not an authorized component of the communication system.

In this paper we propose a location-based beamforming (LBB) scheme that does not require the eavesdropper pass her (instantaneous or statistical) CSI back to the source. Rather, we will assume that some {\em a priori} known location information of the eavesdropper is available at the source. Such a scenario can occur in many circumstances, such as those detailed in \cite{shihao_axiv}. In our scheme, we assume  that \emph{all} of the communication channels are in a Rician fading environment. That is, \emph{all} the channels can vary from pure line-of-sight (LOS) channel to pure Rayleigh channel as the Rician $K$-factors in the channels change.  We also assume that the CSI from the legitimate receiver is \emph{perfectly} known at the source, while the \emph{only} information on the eavesdropper available at the source is her location\footnote{Strictly speaking we must also assume a limit on the size of the eavesdropper's device, as this size-limit in effect places an upper limit of the number of antennas at the eavesdropper that obtain uncorrelated signals. Increasing the number of correlated antennas at the eavesdropper will increase her signal-to-interference-plus-noise-ratio (SINR) \cite{he_biao}. We note that the SINR at the eavesdropper is also limited (ultimately) by the size of the device.}. Our key goal is to determine the beamforming vector at the source that minimizes the secrecy outage probability of the system, given the CSI of the main channel and the eavesdropper's location.

Perhaps the most relevant works to ours are those of \cite{shihao_axiv} and \cite{jemin}. In \cite{shihao_axiv}, the secrecy outage probability of a LBB scheme in Rician wiretap channels was investigated, under the assumptions that the location of the legitimate receiver was available and location of the eavesdropper was available. Different from \cite{shihao_axiv}, our two assumptions are that the location of the eavesdropper is available and the CSI from the legitimate receiver is available. That is, the assumption set we adopt in this work is different. Based on this latter assumption set, we propose a new LBB scheme that minimizes the secrecy outage probability. We note that, the new assumption set we adopt will lead to a reduction in the secrecy outage probability (relative to \cite{shihao_axiv}), but (more importantly) will also enable us to determine the optimal beamforming vector at the source in a more efficient manner.
In \cite{jemin}, the secrecy outage probability was examined in Rician wiretap channels where the source is equipped with a large number of antennas. Different from \cite{jemin}, our proposed scheme applies for an arbitrary number of antennas at the source. Moreover, we introduce a jammer to the system. Our contributions are summarized as follows:

\begin{itemize}
\item We derive a simple expression of the secrecy outage probability when the eavesdropper's location and the CSI of the main channel are known. We highlight that our expression is valid for arbitrary values of SINR and Rician $K$ factors in the main channel and the channel between the source and the eavesdropper.
\item Based on this new expression, we develop a much more efficient search algorithm for the determination of the optimal beamforming scheme that minimizes the secrecy outage probability when the CSI of the main channel and the eavesdropper's location are available at the source. We highlight that our new search algorithm invokes a one-dimensional search, as opposed to the multi-dimensional searches required previously, thereby greatly reducing the computational complexity (important for in-filed deployment).
\item We derive an approximate expression of the secrecy outage probability of the system with the jammer for the special case where the Rician $K$-factor of the jammer-eavesdropper channel is $0$, which provides a computationally efficient way to characterize the secrecy outage probability of the system with the jammer when the jammer-eavesdropper channel is in a pure Rayleigh fading environment.
\item We examine the impact of location uncertainty on the secrecy outage probability, showing how secrecy can still exist when only a noisy estimate of the eavesdropper's location is available at the source.
\end{itemize}

The rest of the paper is organized as follows. Section II describes the system model considered in the paper. In Section III, we detail the proposed LBB scheme in Rician wiretap channels without the jammer. In Section IV, we examine the proposed LBB scheme in Rician wiretap channels with the jammer. Numerical results and related discussions are presented in Section V. Finally, Section VI draws conclusions.

{\em Notations}: Column vectors (matrices) are denoted by boldface lower (upper) case letters. Transpose and conjugate transpose are denoted by $\left(\cdot\right)^T$ and $\left(\cdot\right)^H$, respectively. Complex Gaussian distribution is denoted by $\mathcal{CN}$. An imaginary number is denoted by $j$. A $1\times m$ zero vector is denoted by $\mathbf{0}_{1\times m}$. An $m\times m$ zero matrix and an $m\times m$ identity matrix are denoted by $\mathbf{0}_m$ and $\mathbf{I}_m$, respectively. Statistical expectation and Statistical variance are denoted by $\mathbb{E}$ and $\text{Var}$, respectively. The diagonal elements of a matrix is denoted by $\text{diag}\left[\cdot\right]$. The trace of a matrix is denoted by $\mbox{Tr}\left\{\cdot\right\}$. The absolute value of a scalar is denoted by $|\cdot|$. The Frobenius norm of a vector or a matrix is denoted by $\|\cdot\|$.

\section{System Model}
\begin{figure}[t]
\begin{center}{\includegraphics[width=0.8\columnwidth]{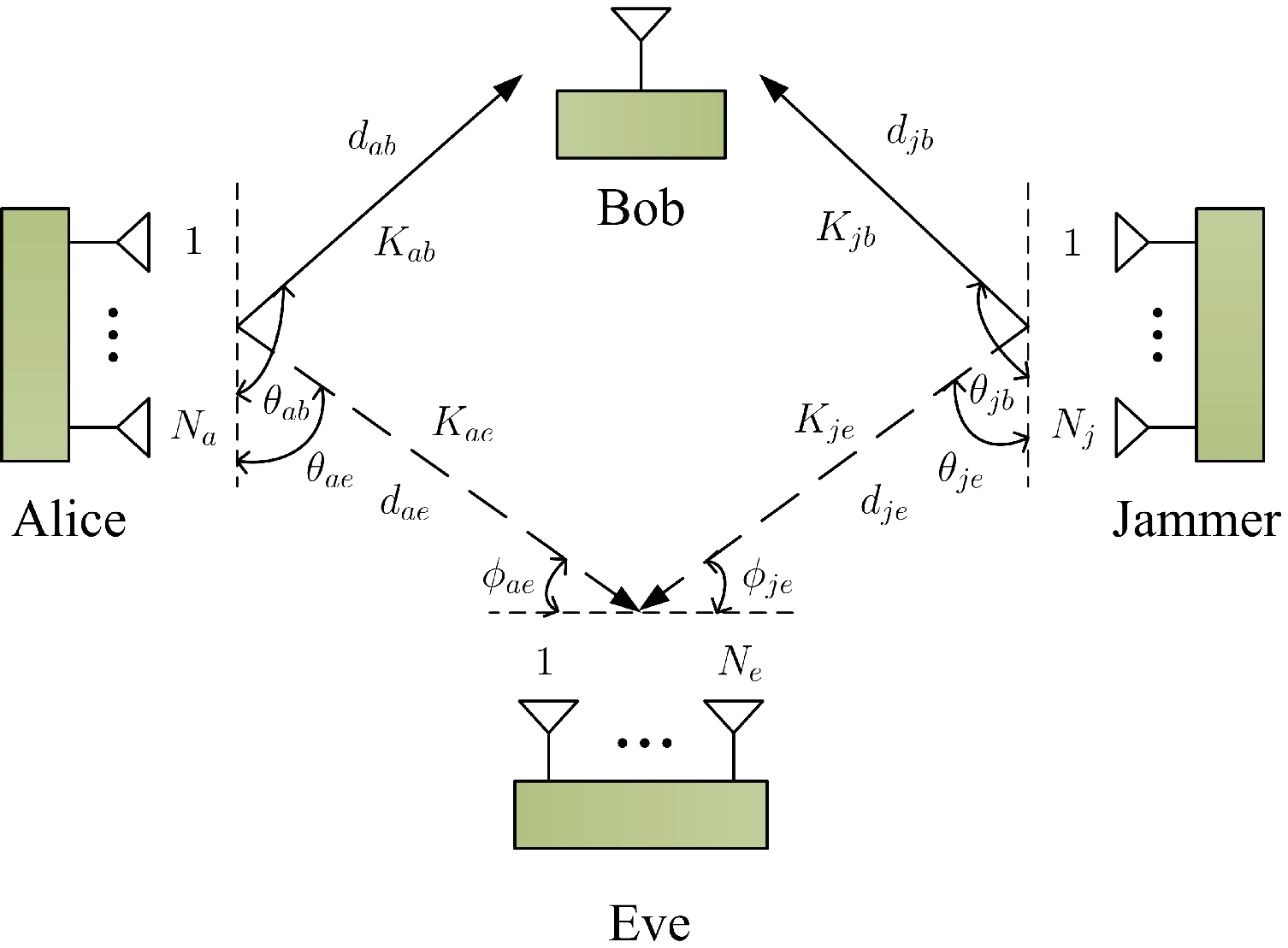}}
\caption{Illustration of our wiretap channel with Rician fading. The Rician fading between all devices is assumed general, covering pure LOS through to pure Rayleigh fading. Real-world-channels lie somewhere in between these extremes. Note this figure serves to define the angles used in the main text and the distances between devices.}\label{system_model}
\end{center}
\end{figure}

We consider a wiretap channel with Rician fading consisting of a source (Alice), a destination (Bob), a Jammer (J), and an eavesdropper (Eve), as shown in Fig. \ref{system_model}. In this channel, Alice communicates with Bob in the presence of Eve. Simultaneously, J transmits the jamming signals to degrade the quality of the received signals at Eve, while maintaining the quality of the received signals at Bob. Alice, J, and Eve are equipped with uniform linear arrays (ULA) with $N_a$, $N_j$ and $N_e$ antennas, respectively, while Bob is equipped with a single antenna.
We adopt the polar coordinate system. As such, the locations of Alice, Bob, J, and Eve are denoted by $\left(0,0\right)$, $\left(d_{ab},\theta_{ab}\right)$, $\left(d_{aj},\theta_{aj}\right)$ and $\left(d_{ae},\theta_{ae}\right)$, respectively. We assume that all the channels are subject to quasi-static independent and identically distributed (i.i.d) Rician fading with different Rician $K$-factors. We  assume that (via  {\em a priori} measurement campaigns) the $K$-factors and path loss exponents of all relevant channels are known, and that
 the CSI of the main channel between Alice and Bob is known to Alice. We also assume that the CSI of the J-Bob channel is known to J. We further assume that Eve's location is available at Alice. We clarify that in practice Alice can obtain Eve's location in a wide range of scenarios \cite{shihao_axiv}. For instance, a military scenario where Eve's location can be determined through some visual surveillance, Eve communicates with other systems (and therefore Alice can determine her location by detecting her signals), and Eve has a fixed known location (e.g., Eve is a base station). To make progress we will first assume Eve's location is known exactly, turning to noisy estimates later in the paper.

We denote $\mathbf{h}_{qb}$, $q\in\left\{a,j\right\}$, as the channel vector from Alice or J to Bob, which is given by
\begin{align}
\label{bob_channel} \mathbf{h}_{qb} = \sqrt{\frac{K_{qb}}{1 + K_{qb}}}\mathbf{h}_{{qb}}^{o}  + \sqrt{\frac{1}{1 + K_{qb}}}\mathbf{h}_{{qb}}^{r},
\end{align}
where $K_{qb}$ denotes the Rician $K$-factor in the channel from Alice ($q=a$) or J ($q=j$) to Bob, $\mathbf{h}_{qb}^{o}$ denotes the LOS component in the channel from Alice or J to Bob, and $\mathbf{h}_{qb}^{r}$ denotes the scattered component in the channel from Alice or J to Bob - the elements of which are assumed to be i.i.d complex Gaussian random variables with zero mean and unit variance. In \eqref{bob_channel}, $\mathbf{h}_{qb}^{o}$ is defined as \cite{Tsai}
\begin{align}
\label{h_o} \mathbf{h}_{qb}^{o} = \begin{bmatrix}
  1, \cdots ,\exp\left(j2\pi\left(N_q-1\right)\right)\delta_q\cos\theta_{qb}
\end{bmatrix},
\end{align}
where $\delta_q$ denotes the constant spacing, in wavelengths between adjacent antennas of the ULA at Alice or J.

We denote $\mathbf{G}_{qe}$ as the channel matrix from Alice or J to Eve, which is given by
\begin{align}
\label{eve_channel} \mathbf{G}_{qe} = \sqrt{\frac{K_{qe}}{1 + K_{qe}}}\mathbf{G}_{{qe}}^{o}  + \sqrt{\frac{1}{1 + K_{qe}}}\mathbf{G}_{{qe}}^{r},
\end{align}
where $K_{qe}$ denotes the Rician $K$-factor in the channel from Alice or J to Eve, $\mathbf{G}_{{qe}}^{o}$ denotes the LOS component in the channel from Alice or J to Eve, and $\mathbf{G}_{{qe}}^{r}$ denotes the scattered component in the channel from Alice or J to Eve -  the elements of which are assumed to be i.i.d complex Gaussian random variables with zero mean and unit variance. In \eqref{eve_channel}, $\mathbf{G}_{{qe}}^{o}$ is expressed as \cite{Taricco},
\begin{align}
\label{G_o} \mathbf{G}_{{qe}}^{o} = \left(\mathbf{r}_{qe}^{o}\right)\mathbf{g}_{qe}^{o},
\end{align}
where $\mathbf{r}_{qe}^{o}$ denotes the array response of Alice or J's transmitted signals at Eve, which is given by
\begin{align}
\label{r_o} \mathbf{r}_{qe}^{o} = \begin{bmatrix}
1,\cdots,\exp\left(-j2\pi\left(N_e-1\right)\delta_e\cos\phi_{qe}\right)
\end{bmatrix},
\end{align}
where $\delta_e$ denotes the constant spacing, in wavelengths, between adjacent antennas of the ULA at Eve, and $\phi_{qe}$ denotes the angle of arrival from Eve to Alice or J (see Fig. \ref{system_model}), and $\mathbf{g}_{qe}^{o}$ denotes the array response at Alice or J, which is given by
\begin{align}
\label{g_o} \mathbf{g}_{qe}^{o} = \begin{bmatrix}
1,\cdots,\exp\left(j2\pi\left(N_q-1\right)\delta_q\cos\theta_{qe}\right)
\end{bmatrix}.
\end{align}
where $\theta_{qe}$ denotes the angle from Alice or J to Eve.

We assume that J transmits the jamming signal to degrade the quality of the received signal at Eve, while maintaining the quality of the received signal at Bob. As such, we design the jamming signal from J as
\begin{align}\label{jamming_signal}
\mathbf{x}_{\text{AN}} = \mathbf{W}_{\text{AN}}\mathbf{t}_{\text{AN}},
\end{align}
where $\mathbf{W}_{\text{AN}}$ is an $N_{j}\times\left(N_j-1\right)$ beamforming matrix used to transmit the jamming signal, and $\mathbf{t}_{\text{AN}}$ is an $\left(N_j-1\right)\times 1$ vector of the jamming signal. In designing $\mathbf{x}_{\text{AN}}$, we choose $\mathbf{W}_{\text{AN}}$ as the orthonormal basis of the null space of $\mathbf{h}_{jb}$. We then choose $\mathbf{t}_{\text{AN}}$ to satisfy $\mathbb{E}\left[\mathbf{\mathbf{t}_{\text{AN}}}\mathbf{t}_{\text{AN}}^H\right] = \frac{1}{N_j-1}\mathbf{I}_{N_j-1}$. Such a design ensures that $\text{Tr}\left\{\mathbf{x}_{\text{AN}}\mathbf{x}_{\text{AN}}^H\right\}=1$.

According to \eqref{bob_channel}--\eqref{jamming_signal}, we express the received signal at Bob as
\begin{align}
\label{y_b} y_b = \sqrt{P_ad_{ab}^{-\eta_{ab}}}\mathbf{h}_{ab}\mathbf{x}_a + n_b,
\end{align}
where $P_a$ denotes the transmit power at Alice, $\eta_{ab}$ denotes the path loss exponent of the main channel (as in \cite{shihao_axiv} all path loss exponents will be a function of the sender-receiver channel, i.e. device locations), and $n_b$ denotes the thermal noise at Bob - which is assumed to be a complex Gaussian random variable with zero mean and variance $\sigma_b^2$, i.e., $n_b\sim\mathcal{CN}\left(0,\sigma_b^2\right)$. In \eqref{y_b}, $\mathbf{x}_a$ is expressed as
\begin{align}
\label{x_a} \mathbf{x}_a = \mathbf{w}t_a,
\end{align}
where $\mathbf{w}$ denotes the $N_a\times 1$ beamforming vector\footnote{We correct a typographical error in \cite{chenxi6} here by setting $\mathbf{w}$ as a row vector.}, and $t_a$ is a scalar, which denotes the information signal transmitted by Alice. We assume that $\|\mathbf{w}\|^2=1$ and $\mathbb{E}\left[|t_a|^2\right]=1$. We then express the received signal at Eve as
\begin{align}
\label{y_e_jammer} \mathbf{y}_{e} = \sqrt{P_ad_{ae}^{-\eta_{ae}}}\mathbf{G}_{ae}\mathbf{x}_a + \sqrt{P_jd_{je}^{-\eta_{je}}}\mathbf{G}_{je}\mathbf{x}_{\text{AN}}+ \mathbf{n}_e,
\end{align}
where $P_j$ denotes the transmit power at J, $d_{je}$ denotes the distance between J and Eve, $\eta_{ae}$ and $\eta_{je}$ denote the path loss exponents in the Alice-Eve channel and J-Eve channel, respectively, and $\mathbf{n}_e$ denotes the thermal noise vector at Eve - the elements of which are assumed to be i.i.d complex Gaussian random variables with zero mean and variance $\sigma_e^2$, i.e., $\mathbf{n}_e\sim\left(\mathbf{0}_{N_e\times 1},\mathbf{I}_{N_e}\right)$.

As such, we express the received SINR at Bob as
\begin{align}
\label{snr_b} \gamma_b = \tilde{\gamma}_{ab}|\mathbf{h}_{ab}\mathbf{w}|^2,
\end{align}
where $\tilde{\gamma}_{ab}={P_ad_{ab}^{-\eta_{ab}}}/{\sigma_b^2}$.

In order to maximize the probability of successful eavesdropping, we assume that Eve applies the minimum mean square error (MMSE) combining to process her received signal. As per the rules of MMSE combining \cite{kim} , we express the instantaneous SINR at Eve as
\begin{align}
\label{gamma_e_jammer} \gamma_{e} = \tilde{\gamma}_{ae}\mathbf{w}^H\mathbf{G}_{ae}^H\mathbf{M}^{-1}\mathbf{G}_{ae}\mathbf{w},
\end{align}
where $\tilde{\gamma}_{ae} = P_ad_{ae}^{-\eta_{ae}}/\sigma_e^2$ and
\begin{align}
\label{K} \mathbf{M} = \frac{\tilde{\gamma}_{je}}{N_j-1}\mathbf{G}_{je}\mathbf{W}_{\text{AN}}\mathbf{W}_{\text{AN}}^H\mathbf{G}_{je}^H+\mathbf{I}_{N_e}
\end{align}
with $\tilde{\gamma}_{je} = P_jd_{je}^{-\eta_{je}}/\sigma_e^2$.

Based on \eqref{snr_b} and \eqref{gamma_e_jammer}, the achievable secrecy rate in the wiretap channel is expressed as \cite{Oggier}
\begin{align}
\label{secrecy_rate} C_s = \left\{\begin{array}{ll}
C_{b}-C_{e}, &\gamma_b>\gamma_e\\0,&\gamma_e\leq\gamma_e,
\end{array}\right.
\end{align}
where $C_b=\log_2\left(1+\gamma_b\right)$ is the capacity of the main channel, and $C_e=\log_2\left(1+\gamma_e\right)$ is the capacity of the Alice-Eve channel.
In this wiretap channel, if $C_s\geq R_s$, where $R_s$ denotes a given secrecy transmission rate, the perfect secrecy is guaranteed. If $C_s< R_s$, information on the transmitted signal is leaked to Eve, and the secrecy is compromised. In order to evaluate the secrecy performance of the wiretap channel in detail, we adopt the secrecy outage probability as the performance metric - defined as the probability that the achievable secrecy rate is less than a given secrecy transmission rate conditioned on $\gamma_b$. Mathematically, this is formulated as
\begin{align}
\label{secrecy_outage_probability} P_{\text{out}}\left(R_s\right) = \mbox{Pr}\left(C_s<R_s|\gamma_b\right).
\end{align}

Our goal is to find the optimal beamforming vector that minimizes the secrecy outage probability. That is, we wish to find
\begin{align}
\label{prob_form} \mathbf{w}^{\ast}=\argmin_{\mathbf{w},\|\mathbf{w}\|^2=1} P_{\text{out}}\left(R_s\right).
\end{align}

\section{Location-Based Beamforming Without Jammer}
In this section we assume that J is not transmitting (i.e., $P_j=0$),
 describing in detail how the optimal beamforming scheme (that minimizes the secrecy outage probability) is obtained through the use of Bob's CSI and Eve's location. We also derive an easy-to-compute expression for the secrecy outage probability when the proposed LBB scheme is applied.

We first re-express $\mathbf{y}_{e}$ in \eqref{y_e_jammer} when J is not transmitting as
\begin{align}
\label{y_e} \mathbf{y}_e = \sqrt{P_ad_{ae}^{-\eta_{ae}}}\mathbf{G}_{ae}\mathbf{x}_a + \mathbf{n}_e.
\end{align}
We then re-express $\gamma_{e}$ in \eqref{gamma_e_jammer} when J is not transmitting as
\begin{align}
\label{snr_e} \gamma_e = \tilde{\gamma}_{ae}\|\mathbf{G}_{ae}\mathbf{w}\|^2.
\end{align}
In order to solve \eqref{prob_form}, we
 present the following proposition.
\begin{proposition}
\label{p1} Given $\tau\in\left[0,1\right]$, the optimal beamforming vector $\mathbf{w}^{\ast}$ that minimizes the secrecy outage probability is a member of the following family of beamformer solutions,
\begin{align}
\label{p1_result} \mathbf{w}\left(\tau\right) = \sqrt{\tau}\mathbf{w}_{1} + \sqrt{1-\tau}\mathbf{w}_{2}.
\end{align}
 Here, $\mathbf{w}_{1} = \frac{\mathbf{\Psi}_{\mathbf{G}_{ae}^{o}}^{\bot}\mathbf{h}_{ab}^H}{\|\mathbf{\Psi}_{\mathbf{G}_{ae}^{o}}^{\bot}\mathbf{h}_{ab}^H\|}$, where $\mathbf{\Psi}_{\mathbf{G}_{ae}^{o}}^{\bot}=\mathbf{I}_{N_a}-\left(\mathbf{G}_{ae}^{o}\right)^H\left(\mathbf{G}_{ae}^{o}\left(\mathbf{G}_{ae}^{o}\right)^H\right)^{-1}\mathbf{G}_{ae}^{o}$; and $\mathbf{w}_{2}= \frac{\mathbf{\Psi}_{\mathbf{G}_{ae}^{o}}\mathbf{h}_{ab}^H}{\|\mathbf{\Psi}_{\mathbf{G}_{ae}^{o}}\mathbf{h}_{ab}^H\|} $, where $\mathbf{\Psi}_{\mathbf{G}_{ae}^{o}}=\left(\mathbf{G}_{ae}^{o}\right)^H\left(\mathbf{G}_{ae}^{o}\left(\mathbf{G}_{ae}^{o}\right)^H\right)^{-1}\mathbf{G}_{ae}^{o}$.
\begin{proof}
Based on \eqref{y_e} and \eqref{snr_e}, we re-express $P_{\text{out}}\left(R_s\right)$ in \eqref{secrecy_outage_probability} when J is not transmitting as
\begin{align}
\label{p1_proof_first} P_{\text{out}}\left(R_s\right) &= \mbox{Pr}\left(C_b-C_e<R_s|\gamma_b\right)\notag\\
&=\mbox{Pr}\left(C_e>C_b-R_s|\gamma_b\right)\notag\\
&=\mbox{Pr}\left(\gamma_e>2^{-R_s}\left(1+\gamma_b\right)-1\right).
\end{align}
According to \eqref{p1_proof_first}, we find that $P_{\text{out}}\left(R_s\right)$ increases as $\gamma_b$ decreases and $\gamma_e$ increases.
Suppose that $\{\mathbf{w}_{1},\mathbf{w}_{2},\mathbf{w}_3,\cdots,\mathbf{w}_{N_a}\}$ denotes an orthonormal basis in the complex space $\mathbb{C}^{N_a}$. As such, any beamforming vector at Alice can be expressed as \cite{gerbracht}
\begin{align}
\label{p1_proof}\mathbf{w} = \lambda_1\mathbf{w}_{1}+\lambda_2\mathbf{w}_{2}+\sum_{l=3}^{N_a}\lambda_l\mathbf{w}_l,
\end{align}
where $\mathbf{\lambda} = [\lambda_1,\lambda_2,\cdots,\lambda_{N_a}]$ are complex and $\|\mathbf{\lambda}\|^2 = 1$. Based on \eqref{snr_b} and \eqref{snr_e},
we first note that both $\gamma_b$ and $\gamma_e$ are functions of $\mathbf{w}$. We then note that $\gamma_b$ decreases when $\lambda_l\neq0$. This is due to the fact that $\mathbf{w}_l$ are orthogonal to the plane spanned by $\left\{\mathbf{w}_1,\mathbf{w}_2\right\}$ and the main channel $\mathbf{h}_{ab}$ lies in this plane. We also find that $\gamma_e$ increases when $\lambda_l\neq0$ unless the Alice-Eve channel $\mathbf{G}_{ae}$ also lies in the plane spanned by $\left\{\mathbf{w}_1,\mathbf{w}_2\right\}$.

Based on the above analysis, we see that $\gamma_b$ decreases and $\gamma_e$ increases when $\lambda_l\neq0$, which leads to the increase in $P_{\text{out}}\left(R_s\right)$. As such, we confirm that we need to set $\lambda_l=0$ in order to minimize the secrecy outage probability, and the optimal beamforming vector has the following structure, given by,
\begin{align}
\label{p1_proof_2} \mathbf{w}\left(\tau\right) = \underbrace{{\sqrt{\tau}\exp\left(j\theta_1\right)}}_{\lambda_1}\mathbf{w}_{1} + \underbrace{{\sqrt{1-\tau}\exp\left(j\theta_2\right)}}_{\lambda_2}\mathbf{w}_{2}.
\end{align}
We note that $\theta_1$ and $\theta_2$  in \eqref{p1_proof_2} are general phases having no impact on $C_s$, thus without loss of generality we can set $\theta_1 = \theta_2 = 0$. Substituting $\theta_1 = \theta_2 =0$ into \eqref{p1_proof_2} we obtain the desired result in \eqref{p1_result}, which completes the proof.
\end{proof}
\end{proposition}

With the aid of {\em Proposition \ref{p1}}, we note that the optimal beamforming vector $\mathbf{w}^{\ast}$ that solves \eqref{prob_form} can be obtained by finding the optimal $\tau^{\ast}$ that minimizes the secrecy outage probability. As such, we re-express \eqref{prob_form} as
\begin{align}
\label{prob_form_2} \tau^{\ast} = \argmin_{0\leq\tau\leq1}P_{\text{out}}\left(R_s\right).
\end{align}
We highlight that {\em Proposition \ref{p1}} provides a far more efficient way of obtaining the optimal beamforming vector $\mathbf{w^{\ast}}$ that solves \eqref{prob_form} compared to an exhaustive search. This is due to the fact that an exhaustive search is performed in the complex space $\mathbb{C}^{N_a}$. Consequently, the computational complexity of the exhaustive search grows exponentially as $N_a$ increases. This is to be compared with our method in {\em Proposition \ref{p1}} which involves a one-dimensional search of $\tau^{\ast}$ only, regardless of the value of $N_a$. We note when Bob is equipped with multiple antennas, and a single-stream transmission from Alice occurs, {\em Proposition \ref{p1}} applies directly. This is due to the fact that placing more antennas at Bob only impacts the received SINR at Bob. In such circumstances the secrecy outage probability decreases. We also note that a similar result of {\em Proposition \ref{p1}} was obtained in \cite{ryu}, which was derived from maximizing the expected achievable rate in cooperative relay networks.

We now present the expression of the secrecy outage probability when $\mathbf{w}\left(\tau\right)$ is adopted as the beamforming vector in the following theorem.

\begin{theorem}
\label{t1} The secrecy outage probability when $\mathbf{w}\left(\tau\right) = \sqrt{\tau}\mathbf{w}_{1} + \sqrt{1-\tau}\mathbf{w}_{2}$ is adopted as the beamforming vector is given by
\begin{align}
\label{t1_result} P_{\text{out}}\left(R_s\right) = 1- \frac{\gamma\left(N_e\hat{m}_{ae},\frac{2^{-R_s}\left(1+\gamma_b\right)-1}{\hat{m}_{ae}^{-1}{\overline{\gamma}}_e}\right)}{\Gamma\left(N_e\hat{m}_{ae}\right)},
\end{align}
where $\gamma\left(\cdot,\cdot\right)$ is the lower incomplete gamma function, defined as \cite[Eq. (8.350)]{table},
\begin{align}
\label{lower_incomplete_function} \gamma\left(\mu,\nu\right) = \int_{0}^{\nu}\exp\left(-t\right)t^{\mu-1}dt,
\end{align}
\begin{align}
\label{m_e} \hat{m}_{ae}=\frac{\left(\hat{K}_{ae}+1\right)^2}{2\hat{K}_{ae}+1},
\end{align}
where $\hat{K}_{ae}=|\mathbf{g}_{ae}^{o}\mathbf{w}\left(\tau\right)|^2K_{ae}$,
\begin{align}
\label{tilde_gamma_e} {\overline{\gamma}}_e = \mathbb{E}\left[\gamma_e\right] =\frac{\left(K_{ae}|\mathbf{g}_{ae}^{o}\mathbf{w}\left(\tau\right)|^2+1\right)\tilde{\gamma}_{ae}}{1+K_{ae}},
\end{align}
and $\Gamma\left(\cdot\right)$ is the Gamma function, defined as \cite[Eq. (8.310)]{table},
\begin{align}
\label{gamma_func} \Gamma\left(z\right) = \int_0^{\infty}\exp\left(-t\right)t^{z-1}dt.
\end{align}
\begin{proof}
We focus on the probability density function (PDF) of $\gamma_e$ when $\mathbf{w}\left(\tau\right)$ is adopted as the beamforming vector,
 which is expressed as \cite{shihao_axiv}
\begin{align}
\label{pdf_snr_e} f_{\gamma_e}\left(x\right) = \left(\frac{\hat{m}_{ae}}{{\overline{\gamma}}_e}\right)^{N_e\hat{m}_{ae}}\frac{x^{N_e\hat{m}_{ae}-1}}{\Gamma\left(N_e\hat{m}_{ae}\right)}\exp\left(-\frac{\hat{m}_{ae}x}{{\overline{\gamma}}_e}\right).
\end{align}
The cumulative distribution function (CDF) of $\gamma_e$ is then obtained as
\begin{align}
\label{CDF_snr_e} F_{\gamma_e}\left(x\right) = \frac{\gamma\left(N_e\hat{m}_{ae},\frac{\hat{m}_{ae}x}{{\overline{\gamma}}_e}\right)}{\Gamma\left(N_e\hat{m}_{ae}\right)}.
\end{align}
We then re-express $P_{\text{out}}\left(R_s\right)$ in \eqref{p1_proof_first} as
\begin{align}
\label{secrecy_outage_probability_2} P_{\text{out}}\left(R_s\right)
&=1-F_{\gamma_e}\left(2^{-R_s}\left(1+\gamma_b\right)-1\right).
\end{align}
Substituting \eqref{CDF_snr_e} into \eqref{secrecy_outage_probability_2}, we obtain the desired result in {\em Theorem \ref{t1}}. The proof is completed.
\end{proof}
\end{theorem}

Note, in {\em Theorem \ref{t1}} Eve's location is explicitly expressed in the expressions for $\hat{m}_{ae}$, $\hat{K}_{ae}$, and ${\overline{\gamma}}_e$. Note also, that our derived expression is valid for arbitrary values of average SINRs and Rician $K$-factors in the main channel and the Alice-Eve channel. In the following, we detail how the optimal $\tau^{\ast}$ that minimizes $P_{\text{out}}\left(R_s\right)$ can be obtained per block by applying {\bf Algorithm \ref{a1}}.

\begin{algorithm}[htb]
\caption{Algorithm to determine $\tau^{\ast}$ per block when J is not transmitting}
\label{a1}
\begin{algorithmic}[1]
\REQUIRE $\mathbf{h}_{ab}$
\ENSURE $\tau^{\ast}$.
\STATE Calculate $\mathbf{w}_1$ and $\mathbf{w}_2$.
\FOR{every $\tau\in\left[0,1\right]$ with step size $\delta_t$}
\STATE Calculate $\mathbf{w}\left(\tau\right)$ using \eqref{p1_result}.
\STATE Calculate $P_{\text{out}}\left(R_s\right)$ using \eqref{t1_result}.
\ENDFOR
\STATE Choose $\tau^{\ast}$ as the value of $\tau$ that achieves the minimum $P_{\text{out}}\left(R_s\right)$.
\end{algorithmic}
\end{algorithm}
We now evaluate the computational demands of \textbf{Algorithm \ref{a1}}. For a given $N_a$, $N_e$, and $\delta_t$, \textbf{Algorithm \ref{a1}} requires $2\delta_t^{-1}$ gamma function calculations. We note that the complexity for  the gamma function calculation is $O\left(n^{5/2}\left(\log n\right)^2\right)$ \cite{borwein},  where $n$ denotes the number of digits used. For anticipated values of $N_a$, $N_e$,  $\delta_t = 10^{-2}$, and assuming 64-bit processing,  the number of floating-point operations for \textbf{Algorithm \ref{a1}} is of order $10^{6}$ (note, $\delta_t =10^{-2}$ leads to a negligible error of 1 part in  $10^{4}$ compared to the true minimum secrecy outage probability). Assuming 4 floating-point operations per cycle, $10^{6}$ operations can be completed on a single-core $64$-bit $2.5$ GHz microprocessor  within $1$ ms.
 As such, \textbf{Algorithm \ref{a1}} can be performed in real-time with negligible latency impact\footnote{The computation time of \textbf{Algorithm \ref{a1}} on MATLAB is less than $0.25$ s on a quad-core $64$-bit $3.3$ GHz Intel i$5$-$2500$ microprocessor. Based on the conversion factor from MATLAB to embedded C++ firmware, we estimate the per block latency to be less than $0.5$ ms \cite{andrew}, while the coherence time of the channel is  hundreds of milliseconds for stationary nodes \cite{perahia}.}.  On the other hand, an exhaustive search of the optimal beamforming vector $\mathbf{w}^{\ast}$ that minimizes $P_{\text{out}}\left(R_s\right)$ requires $2\left(\delta_t^{-1}\right)^{2N_a}$ gamma function calculations. We note that  $2\left(\delta_t^{-1}\right)^{2N_a}$ gamma function calculations require of order $10^{12}$ floating-point operations for $N_a = 2$, and  $\delta_t=10^{-2}$. Moreover, the number of floating-point operations for an exhaustive search  grows exponentially with $N_a$. As such, an exhaustive search is simply not practical in real-world deployments.

We point out that $\phi_{ae}$ disappears in the expression for the secrecy outage probability in {\em Theorem \ref{t1}}. As an aside, it is perhaps interesting to show why this is so. To this end, we re-express $\gamma_e$ in \eqref{snr_e} as
\begin{align}
\label{snr_e_re} \gamma_e = \tilde{\gamma}_{ae}\sum_{i=1}^{N_e}|\mathbf{g}_{ae,i}\mathbf{w}\left(\tau\right)|^2,
\end{align}
where $\mathbf{g}_{ae,i}$ is the $1\times N_a$ channel vector between Alice and $i$-th Eve's antenna, given by
\begin{align}
\label{g_i} \mathbf{g}_{ae,i} = \sqrt{\frac{K_{ae}}{1+K_{ae}}}r_{ae,i}^{o}\mathbf{g}_{ae}^{o} + \sqrt{\frac{1}{1+K_{ae}}}\mathbf{g}_{ae,i}^{r},
\end{align}
where $r_{ae,i}^{o}$ is the $i$-th element of $\mathbf{r}_{ae}^{o}$, given by $r_{ae,i}^{o}= \exp\left(-j2\pi\left(i-1\right)\delta_{e}\cos\phi_{ae}\right)$ and $\mathbf{g}_{ae,i}^{r}$ is the $i$-th row of $\mathbf{G}_{ae}^{r}$. Based on \eqref{g_i}, we express $\mathbf{g}_{ae,i}\mathbf{w}\left(\tau\right)$ as
\begin{align}
\label{g_i_2} \mathbf{g}_{ae,i}\mathbf{w}\left(\tau\right) = &\sqrt{\frac{K_{ae}}{1+K_{ae}}}r_{ae,i}^{o}\mathbf{g}_{ae}^{o}\mathbf{w}\left(\tau\right) \notag\\
&\hspace{2cm}+ \sqrt{\frac{1}{1+K_{ae}}}\mathbf{g}_{ae,i}^{r}\mathbf{w}\left(\tau\right).
\end{align}
We note that $|r_{ae,i}^{o}\mathbf{g}_{ae}^{o}\mathbf{w}\left(\tau\right)|^2 = |\mathbf{g}_{ae}^{o}\mathbf{w}\left(\tau\right)|^2$ for any $r_{ae,i}^{o}$. As such, we confirm that $\mathbf{r}_{ae}^{o}$ has no impact on the secrecy outage probability. This reveals that our analysis is also applicable for antennas arrays other than ULA at Eve, since different antenna arrays at Eve only impact $\mathbf{r}_{ae}^{o}$. In addition, we confirm that our analysis is also applicable for antenna arrays other than ULA at Alice and Bob, respectively. This is because we assume that the CSI of the Alice-Bob channel is known to Alice.
\section{Location-Based Beamforming With Jammer}
 In this section we examine the case when J is transmitting (i.e., $P_j > 0$). We shall see of course that a jammer assists the performance. We will also see that, in principal, a modified (more complex) \textbf{Algorithm \ref{a1}} can be used to determine the optimal beamformer in the presence of the jammer. However, we will also see that the previous beamforming solution derived directly from \textbf{Algorithm \ref{a1}}, when used in the presence of a jammer, leads to a performance that is very close to optimal when the number of antenna at Alice is greater than two. This means that in practice the beamforming solution derived from \textbf{Algorithm \ref{a1}} will actually suffice in most circumstances.

To make progress, we present the following proposition.
\begin{proposition}
\label{p2} Given $\tau\in\left[0,1\right]$, the optimal beamforming vector $\mathbf{w}^{\ast}$ that minimizes the secrecy outage probability of Rician wiretap channels with a jammer is also a member of the following family of beamformer solutions,
\begin{align}
\label{p2_result} \mathbf{w}\left(\tau\right) = \sqrt{\tau}\mathbf{w}_{1} + \sqrt{1-\tau}\mathbf{w}_{2}.
\end{align}
\begin{proof}
Let $\mathbf{R} = \frac{\tilde{\gamma}_{je}}{N_j-1}\mathbf{G}_{je}\mathbf{W}_{\text{AN}}\mathbf{W}_{\text{AN}}^H\mathbf{G}_{je}^H$,
the eigenvalue decomposition of $\mathbf{R}$ is given by
$\mathbf{R} = \mathbf{U}^H\mathbf{\Lambda}\mathbf{U}$,
where $\mathbf{U}$ is an $N_e\times N_e$ unitary matrix, and $\mathbf{\Lambda} = \text{diag}\left[\Lambda_1,\cdots,\Lambda_{N_e}\right]$, and where $\Lambda_1,\cdots,\Lambda_{N_e}$ are eigenvalues of $\mathbf{R}$.
Based on $\mathbf{R}$ and $\mathbf{\Lambda}$, we re-express the instantaneous SINR at Eve in \eqref{gamma_e_jammer} as
\begin{align}
\label{gamma_e_jammer_2} \gamma_{e} = & \tilde{\gamma}_{ae}\mathbf{w}^H\mathbf{G}_{ae}^H\mathbf{U}^H\left(\mathbf{\Lambda}+\mathbf{I}_{N_e}\right)^{-1}\mathbf{U}\mathbf{G}_{ae}\mathbf{w}\notag\\
=&\tilde{\gamma}_{ae}\begin{bmatrix}\mu_1^H,\cdots,\mu_{N_e}^H\end{bmatrix}
\begin{bmatrix}
  \frac{1}{\Lambda_1+1},&\cdots,&0\\
  \vdots&\ddots&\vdots\\
  0,&\cdots,&\frac{1}{\Lambda_{N_e}+1}
\end{bmatrix}
\begin{bmatrix}
{\mu_1}\\\vdots\\{\mu_{N_e}}
\end{bmatrix}\notag\\
=&\sum_{i=1}^{N_e} \frac{|\mu_i|^2}{\Lambda_i+1},
\end{align}
where $u_i$ denotes the $i$-th element of $\mathbf{U}\mathbf{G}_{ae}\mathbf{w}$. Suppose $\nu_i$ is the $i$-th element of $\mathbf{G}_{ae}\mathbf{w}$, we express the instantaneous SINR at Eve when J is not transmitting as
\begin{align}
\label{gamma_e_jammer_no} \gamma_e = & \tilde{\gamma}_{ae}\mathbf{w}^H\mathbf{G}_{ae}^H\mathbf{G}_{ae}\mathbf{w}\notag\\
=&\tilde{\gamma}_{ae}\begin{bmatrix}\nu_1^H,\cdots,\nu_{N_e}^H\end{bmatrix}
\begin{bmatrix}\nu_1\\\vdots\\\nu_{N_e}\end{bmatrix} \notag\\
=&\tilde{\gamma}_{ae}\sum_{i=1}^{N_e}|\nu_i|^2.
\end{align}
We note that $\nu_i$ and $\mu_i$ have the same PDF due to the fact that $\mathbf{U}$ is a unitary matrix. We re-express $P_{\text{out}}\left(R_s\right)$ in \eqref{secrecy_outage_probability} as
\begin{align}
\label{secrecy_outage_probability_4}\hspace{-0.2cm} P_{\text{out}}\left(R_s\right) =& \mbox{Pr}\left(C_b-C_{e}<R_s|\gamma_b\right)\notag\\
=&\mbox{Pr}\left(C_{e}>C_b-R_s|\gamma_b\right)\notag\\
=&\mbox{Pr}\left(\gamma_{e}>2^{-R_s}\left(1+\gamma_b\right)-1\right)\notag\\
=&\mbox{Pr}\left(\tilde{\gamma}_{ae}\sum_{i=1}^{N_e} \frac{|\mu_i|^2}{\Lambda_i+1}>2^{-R_s}\left(1+\gamma_b\right)-1\right)\notag\\
=&\mbox{Pr}\left(\!\tilde{\gamma}_{ae}\sum_{i=1}^{N_e} {|\mu_i|^2}\!>\!\frac{1}{k}{\left(2^{-R_s}\left(1+\gamma_b\right)-1\right)}\!\right),
\end{align}
where
$
k = \frac{\sum_{i=1}^{N_e} \frac{|\mu_i|^2}{\Lambda_i+1}}{\sum_{i=1}^{N_e}|\mu_i|^2}.
$
We then re-express $P_{\text{out}}\left(R_s\right)$ in
\eqref{secrecy_outage_probability} when J is not transmitting as
\begin{align}
\label{secrecy_outage_probability_6} P_{\text{out}}\left(R_s\right)=\mbox{Pr}\left(\tilde{\gamma}_{ae}\sum_{i=1}^{N_e} {|\nu_i|^2}>2^{-R_s}\left(1+\gamma_b\right)-1\right).
\end{align}
Since $\nu_i$ and $\mu_i$ have the same PDF, we re-write \eqref{secrecy_outage_probability_4} as
\begin{align}
\label{secrecy_outage_probability_7} \hspace{-0.2cm}P_{\text{out}}\left(R_s\right) =& \mbox{Pr}\left(\!\tilde{\gamma}_{ae}\sum_{i=1}^{N_e} {|\mu_i|^2}\!>\!\frac{1}{k}{\left(2^{-R_s}\left(1+\gamma_b\right)-1\right)}\!\right)\notag\\
=&\mbox{Pr}\left(\!\tilde{\gamma}_{ae}\sum_{i=1}^{N_e} |{\nu_i|^2}\!>\!\frac{1}{k}{\left(2^{-R_s}\left(1+\gamma_b\right)-1\right)}\!\right).
\end{align}
Observing \eqref{secrecy_outage_probability_6} and \eqref{secrecy_outage_probability_7}, we find that the only difference between $P_{\text{out}}\left(R_s\right)$ when J is not transmitting and $P_{\text{out}}\left(R_s\right)$ when J is transmitting is the factor $k^{-1}$.
Therefore, our analysis in {\em Proposition \ref{p1}}, which is suitable for Rician wiretap channels without the jammer, still holds for channels with the jammer.
According to {\em Proposition \ref{p1}}, the optimal beamforming vector that minimizes $P_{\text{out}}\left(R_s\right)$ when J is not transmitting is a member of the family of beamformer solutions, given by $\mathbf{w}\left(\tau\right) = \sqrt{\tau}\mathbf{w}_1 + \sqrt{1-\tau}\mathbf{w}_2$. As such, we obtain that the optimal beamforming vector that minimizes $P_{\text{out}}\left(R_s\right)$ when J is transmitting is also a member of such a family of beamformer solutions. We note that the optimal value of $\tau$ that minimizes $P_{\text{out}}\left(R_s\right)$ when J is transmitting is different from the optimal $\tau^{\ast}$ that minimizes $P_{\text{out}}\left(R_s\right)$ when J is not transmitting due to the factor $k^{-1}$.
The proof is completed.
\end{proof}
\end{proposition}

According to {\em Proposition \ref{p2}}, we note that the optimal $\mathbf{w}^{\ast}$ at Alice that minimizes $P_{\text{out}}\left(R_s\right)$ can be obtained by determining the optimal $\tau^{\ast}_{j}$ that minimizes $P_{\text{out}}\left(R_s\right)$ when J is transmitting. As such, we re-express \eqref{prob_form} as
\begin{align}
\label{prob_form_4} \tau^{\ast}_j = \argmin_{0\leq\tau\leq1}P_{\text{out}}\left(R_s\right).
\end{align}
We note that the analytical form of $P_{\text{out}}\left(R_s\right)$ for general $K_{je}$ is mathematically intractable since we cannot obtain the closed-form expression for the PDF of $\gamma_e$. As such, in order to obtain the optimal $\tau^{\ast}_j$ that minimizes $P_{\text{out}}\left(R_s\right)$, we apply a modified \textbf{Algorithm \ref{a1}}, in which we numerically calculate $P_{\text{out}}\left(R_s\right)$. Specifically, we first generate $N$ realizations of $\mathbf{G}_{ae}^{r}$ and $\mathbf{G}_{je}^{r}$, we then calculate $C_{s}$ using \eqref{secrecy_rate} for every $\mathbf{G}_{ae}^{r}$ and $\mathbf{G}_{je}^{r}$. Finally, we calculate $P_{\text{out}}\left(R_s\right)$ using \eqref{secrecy_outage_probability}.
For the same level of performance we find the modified algorithm costs approximately 10 times more computational time relative to \textbf{Algorithm \ref{a1}} - and therefore is still viable in real-world deployments. However, as we discuss later, we shall see that in practice the solution provided directly by \textbf{Algorithm \ref{a1}} will actually suffice in most circumstances  - even when the jammer is present.

Moreover, we find that an approximate expression of $P_{\text{out}}\left(R_s\right)$ for the special case where $K_{je} = 0$ is obtainable. We note that such a special case is practical in scenarios where the J-Eve channel is completely blocked (e.g. Eve is in hiding) by buildings.
In order to examine the approximate expression of $P_{\text{out}}\left(R_s\right)$ for the special case where $K_{je} = 0$, we first introduce several new notations as follows:
\begin{align}\label{varphi_l}
\varphi_l = \frac{1}{\left(1+K_{ae}\right)^l}\sum_{m=0}^{l}{l\choose m}\frac{\left(K_{ae}|\mathbf{G}_{ae}^{o}\mathbf{w}\left(\tau\right)|^2\right)^m}{\left(N_e\right)_m},l\in\left\{1,2\right\}
\end{align}
with $\left(N_e\right)_m = \frac{\Gamma\left(N_e+m\right)}{\Gamma\left(N_e\right)}$, and
\begin{align}
\label{vartheta_l}\vartheta_l =& \frac{l\exp\left(\frac{1}{\kappa}\right)}{\kappa^{N_j-1}}\sum_{p = 0}^{N_e-1}\rho_p
\sum_{t=0}^{l-1+p}{l-1+p\choose t}\left(-\frac{1}{\kappa}\right)^{l-1+p-t}\notag\\
&\times\Gamma\left(t-N_j+2,\frac{1}{\kappa}\right),
\end{align}
respectively. In \eqref{vartheta_l}, $\kappa = \frac{\tilde{\gamma}_{je}}{N_j-1}$,
\begin{align}
\label{rho} \rho_p = \kappa^p\sum_{q=\max\left(0,p-N_j+1\right)}^{p}{N_j-1\choose p-q}\frac{1}{q!\kappa^q},
\end{align}
and $\Gamma\left(\cdot,\cdot\right)$ is the upper incomplete Gamma function, defined as \cite[Eq. (8.350)]{table}
\begin{align}
\label{upper_gamma} \Gamma\left(\mu,\nu\right) = \int_{\nu}^{\infty}\exp\left(-t\right)t^{\mu-1}dt.
\end{align}
\begin{theorem}
\label{t2} When $K_{je} = 0$, the approximate secrecy outage probability of Rician wiretap channels with a jammer is
\begin{align}
\label{t2_result} P_{\text{out}}\left(R_s\right)= 1-\frac{\gamma\left(\alpha,\frac{2^{-R_s}\left(1+\gamma_b\right)-1}{\beta}\right)}{\Gamma\left(\alpha\right)},
\end{align}
where
\begin{align}
  \label{t2_result_2} \alpha = \frac{\varphi_1^2\vartheta_1^2}{\varphi_2\vartheta_2-\varphi_1^2\vartheta_1^2},
\end{align}
and
\begin{align}
\label{t2_result_3} \beta = \tilde{\gamma}_{ae}\frac{\left(\varphi_2\vartheta_2-\varphi_1^2\vartheta_1^2\right)}{\varphi_1\vartheta_1}.
\end{align}
\begin{proof}
See Appendix \ref{App_t2}.
\end{proof}
\end{theorem}

We highlight that the approximate expression of the secrecy outage probability in \eqref{t2_result} is valid for arbitrary values of average SINRs and Rician $K$-factors in the main channel and the Alice-Eve channel.

\section{Numerical Results}
In this section, we present numerical results to validate our analysis. Specifically, we first demonstrate the effectiveness of the proposed LBB scheme in Rician wiretap channels where only Alice, Bob, and Eve are involved. We then demonstrate the effectiveness of the scheme when J is transmitting. Finally, we examine the impact of Eve's location uncertainty on secrecy performance of the scheme. Throughout this section, we assume that all the channels have the same path loss exponent, i.e., $\eta_{ab} = \eta_{ae} =\eta_{je} = 4$.

\begin{figure}[t!]
\begin{center}{\includegraphics[height=2.8in,width = 3.0in]{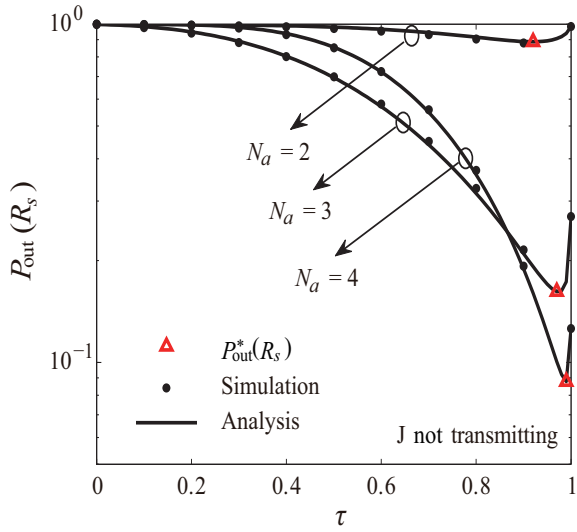}}
\caption{$P_{\text{out}}\left(R_s\right)$ versus $\tau$ for different values of $N_a$ with $N_e=2$, $K_{ab}=10$ dB, $K_{ae}=5$ dB, $\tilde{\gamma}_{ab}=\tilde{\gamma}_{ae}= 10$ dB, $\theta_{ab}=\pi/3$, $\theta_{ae} = \pi/4$, and $R_s=1$ bits/s/Hz. J is not transmitting here.}\label{fig_side_a}
\end{center}
\end{figure}

\begin{figure}[t!]
\begin{center}{\includegraphics[height=2.8in,width = 3.0in]{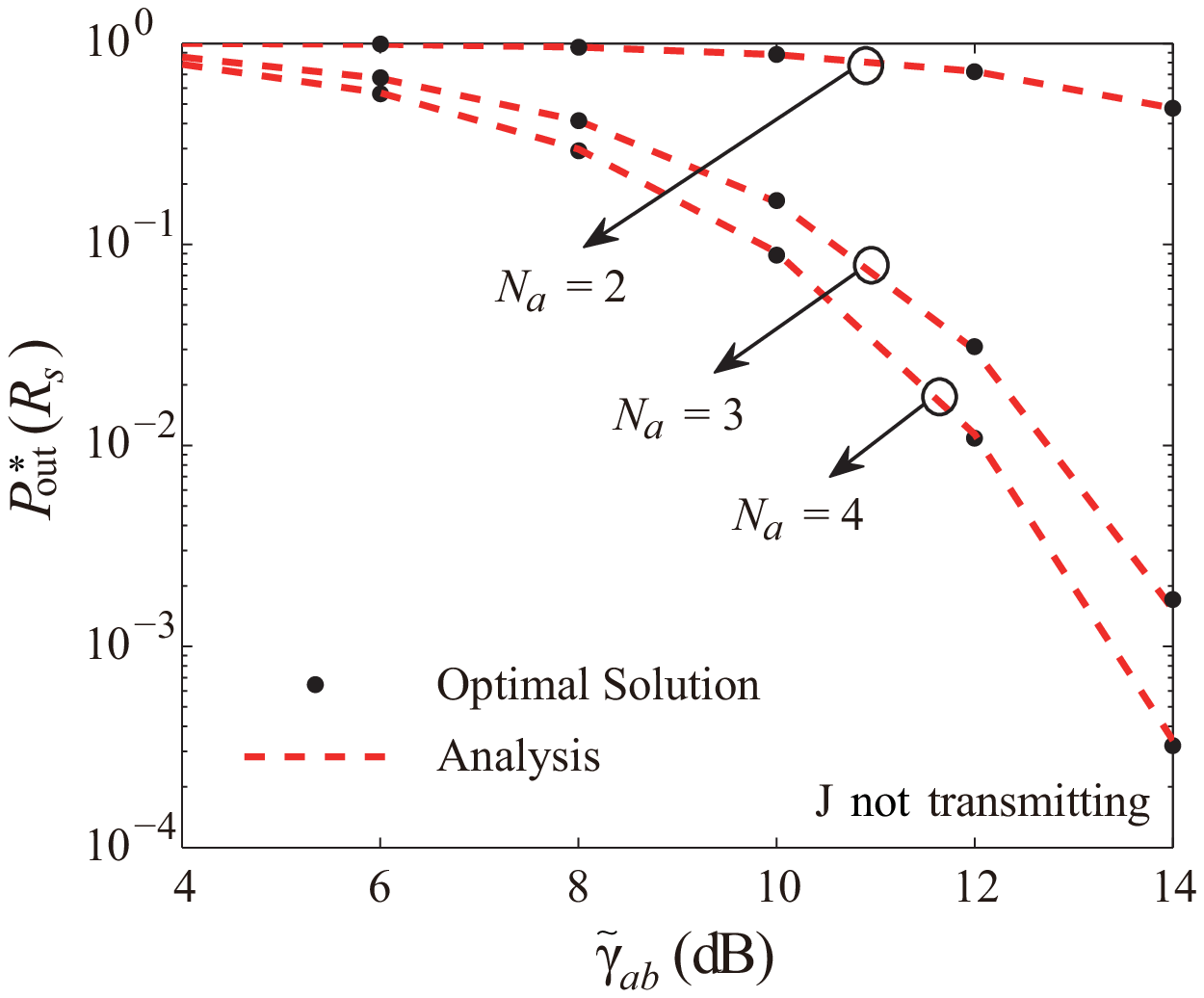}}
\caption{$P_{\text{out}}^{\ast}\left(R_s\right)$ versus $\tilde{\gamma}_{ab}$ for different values of $N_a$ with $N_e=2$, $K_{ab}=10$ dB, $K_{ae}=5$ dB, $\tilde{\gamma}_{ae}= 10$ dB, $\theta_{ab}=\pi/3$, $\theta_{ae} = \pi/4$, and $R_s=1$ bits/s/Hz.  J is not transmitting here.}\label{fig_side_b}
\end{center}
\end{figure}

\begin{figure}[t!]
\begin{center}{\includegraphics[height=2.8in,width = 3.0in]{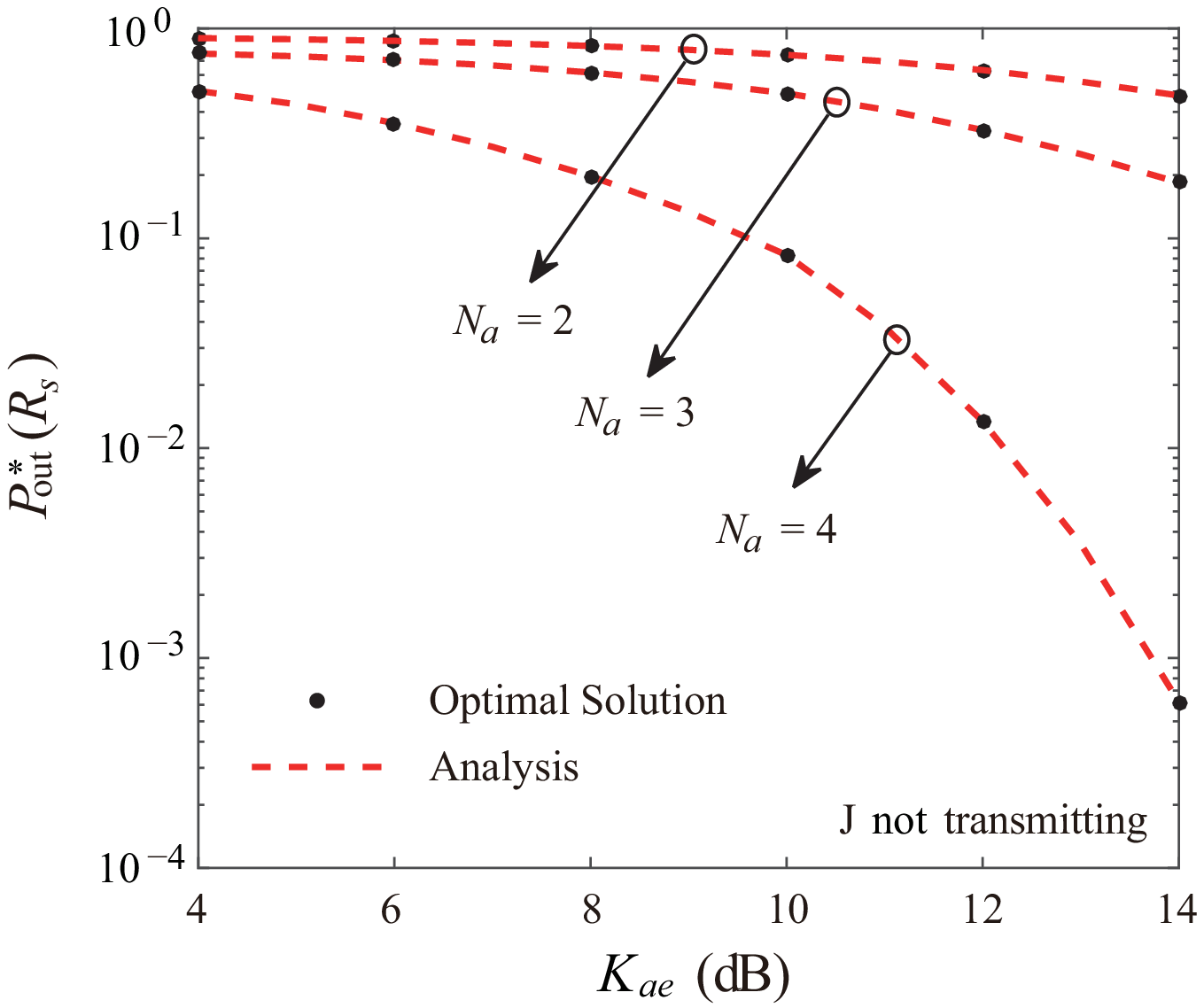}}
\caption{$P_{\text{out}}^{\ast}\left(R_s\right)$ versus $K_{ae}$ for different values of $N_a$ with $N_e=2$, $K_{ab}=10$ dB, $\tilde{\gamma}_{ae}= 10$ dB, $\theta_{ab}=\pi/3$, $\theta_{ae} = \pi/4$, and $R_s=1$ bits/s/Hz.  J is not transmitting here.}\label{fig_side_4}
\end{center}
\end{figure}

We first examine the effectiveness of the scheme when J is not transmitting in Figs. \ref{fig_side_a}--\ref{fig_side_4}.
In Fig. \ref{fig_side_a}, we plot $P_{\text{out}}\left(R_s\right)$ versus $\tau$ for different values of $N_a$ with $N_e=2$, $K_{ab}=10$~dB, $K_{ae}=5$~dB, $\tilde{\gamma}_{ab}=\tilde{\gamma}_{ae}= 10$~dB, $\theta_{ab}=\pi/3$, $\theta_{ae} = \pi/4$, and $R_s=1$ bits/s/Hz. We first observe that the analytical curves, generated from {\em Proposition \ref{p1}} and {\em Theorem \ref{t1}}, precisely match the simulation points marked by black dots, thereby demonstrating the correctness of our analysis for $P_{\text{out}}\left(R_s\right)$ in {\em Proposition \ref{p1}} and {\em Theorem \ref{t1}}. Second, we see that there exists a unique $\tau^{\ast}$ that minimizes $P_{\text{out}}\left(R_s\right)$ for each $N_a$. Third, we see that the minimum $P_{\text{out}}\left(R_s\right)$, denoted by $P_{\text{out}}^{\ast}\left(R_s\right)$, decreases significantly as $N_a$ increases. Furthermore, we observe that the optimal $\tau^{\ast}$ that achieves $P_{\text{out}}^{\ast}\left(R_s\right)$ approaches $1$ as $N_a$ increases. This reveals that the optimal beamforming vector $\mathbf{w}^{\ast}$ that minimizes $P_{\text{out}}\left(R_s\right)$ approaches $\mathbf{w}_{1}$ as $N_a$ increases.

In Fig. \ref{fig_side_b}, we plot $P_{\text{out}}^{\ast}\left(R_s\right)$ versus $\tilde{\gamma}_{ab}$ for different values of $N_a$. In this figure, we have adopted the same system configurations as those in Fig. \ref{fig_side_a}. The analytical curves, represented by red dashed lines, are generated from {\em Proposition \ref{p1}} and {\em Theorem \ref{t1}} with the optimal $\tau^{\ast}$ which minimizes $P_{\text{out}}\left(R_s\right)$ being selected for different values of $N_a$. The optimal beamformer solutions, represented by `$\bullet$' symbols, are obtained from minimizing $P_{\text{out}}\left(R_s\right)$ via an exhaustive search (i.e., a full multi-dimensional search) for different values of $N_a$. We first see that the minimum secrecy outage probability $P_{\text{out}}^{\ast}\left(R_s\right)$ achieved by our scheme is almost the same as the optimal beamformer solution found via exhaustive search. This shows the optimality of our scheme. Second, we see that $P_{\text{out}}^{\ast}\left(R_s\right)$ decreases significantly as $N_a$ increases. This reveals that adding extra transmit antennas at Alice improves the secrecy of the adopted system. We further see that $P_{\text{out}}^{\ast}\left(R_s\right)$ monotonically decreases as $\tilde{\gamma}_{ab}$ increase. This reveals that the secrecy outage probability reduces when Alice uses a higher power to transmit. Moreover, we note that the secrecy outage probability achieved by our proposed scheme outperforms that of solution from \cite{shihao_axiv}. For instance, the secrecy outage probability of our proposed scheme is almost three orders of magnitude less than that of solution from \cite{shihao_axiv} when $N_a = 3$ and $\tilde{\gamma}_{ab} = 14$ dB. This is due to the fact that we determine the optimal beamforming vector that minimizes the secrecy outage probability utilizing the CSI of the main channel and Eve's location, while the solution of \cite{shihao_axiv} was determined using Bob's location and Eve's location.

In Fig. \ref{fig_side_4}, we plot $P_{\text{out}}^{\ast}\left(R_s\right)$ versus $K_{ae}$ for different values of $N_a$. As in Fig. \ref{fig_side_b}, the analytical curves are generated from {\em Proposition \ref{p1}} and {\em Theorem \ref{t1}} with the optimal $\tau^{\ast}$ being selected for a given $N_a$.  Again, we see that the analytical curves match the simulation points. We also see that as expected $P_{\text{out}}^{\ast}\left(R_s\right)$ decreases as $K_{ae}$ increases.

\begin{figure}[t!]
\begin{center}{\includegraphics[height=2.8in,width = 3.0in]{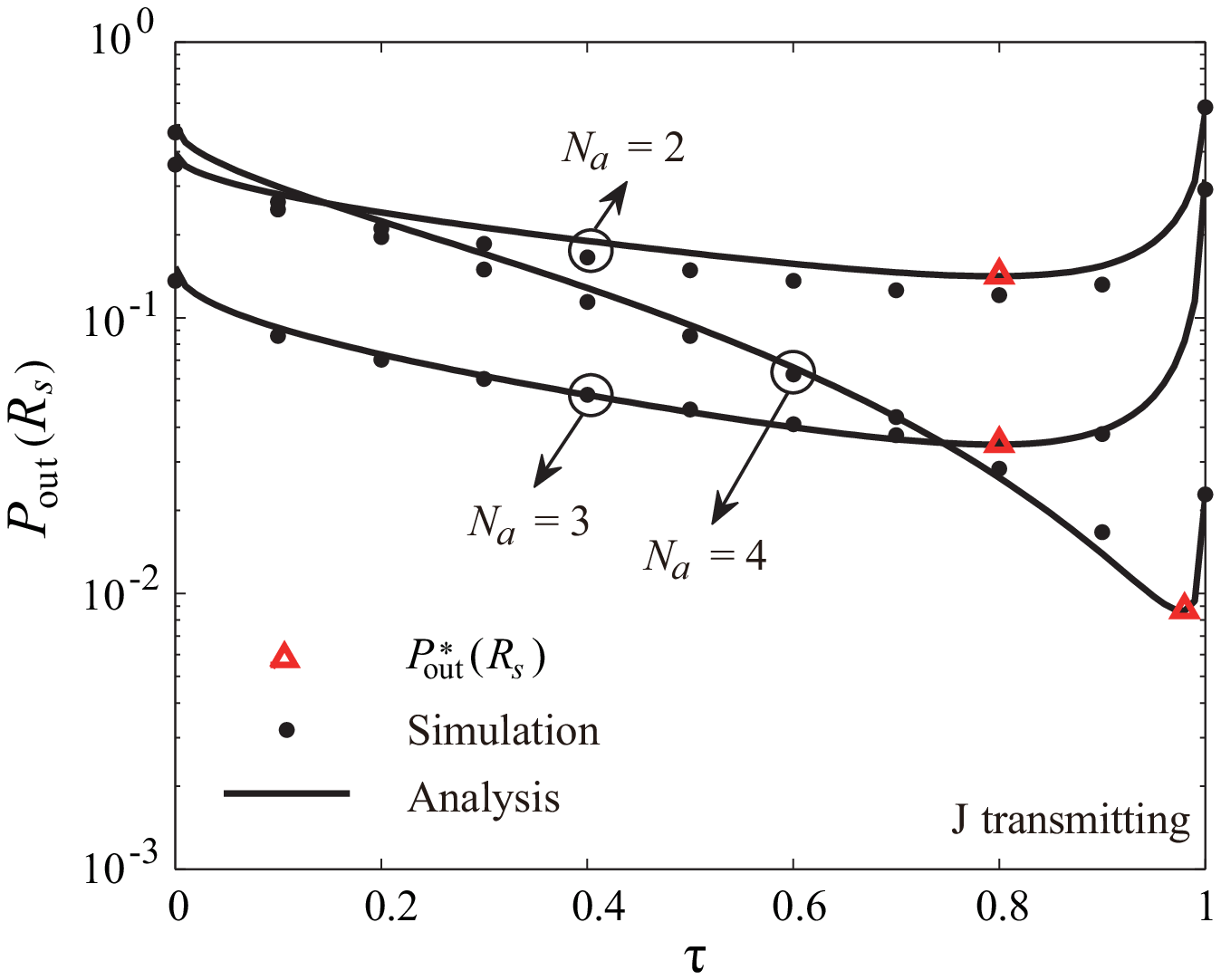}}
\caption{$P_{\text{out}}\left(R_s\right)$ versus $\tau$ for different values of $N_a$ with $N_e=2$, $N_j = 4$, $K_{ab}=10$ dB, $K_{ae}=5$ dB, $K_{je} = 0$, $\tilde{\gamma}_{ab}=\tilde{\gamma}_{ae} = \tilde{\gamma}_{je}= 10$ dB, $\theta_{ab}=\pi/3$, $\theta_{ae} = \pi/4$, and $R_s=1$ bits/s/Hz. J is transmitting here.}\label{fig_side_c}
\end{center}
\end{figure}

\begin{figure}[t!]
\begin{center}{\includegraphics[height=2.8in,width = 3.0in]{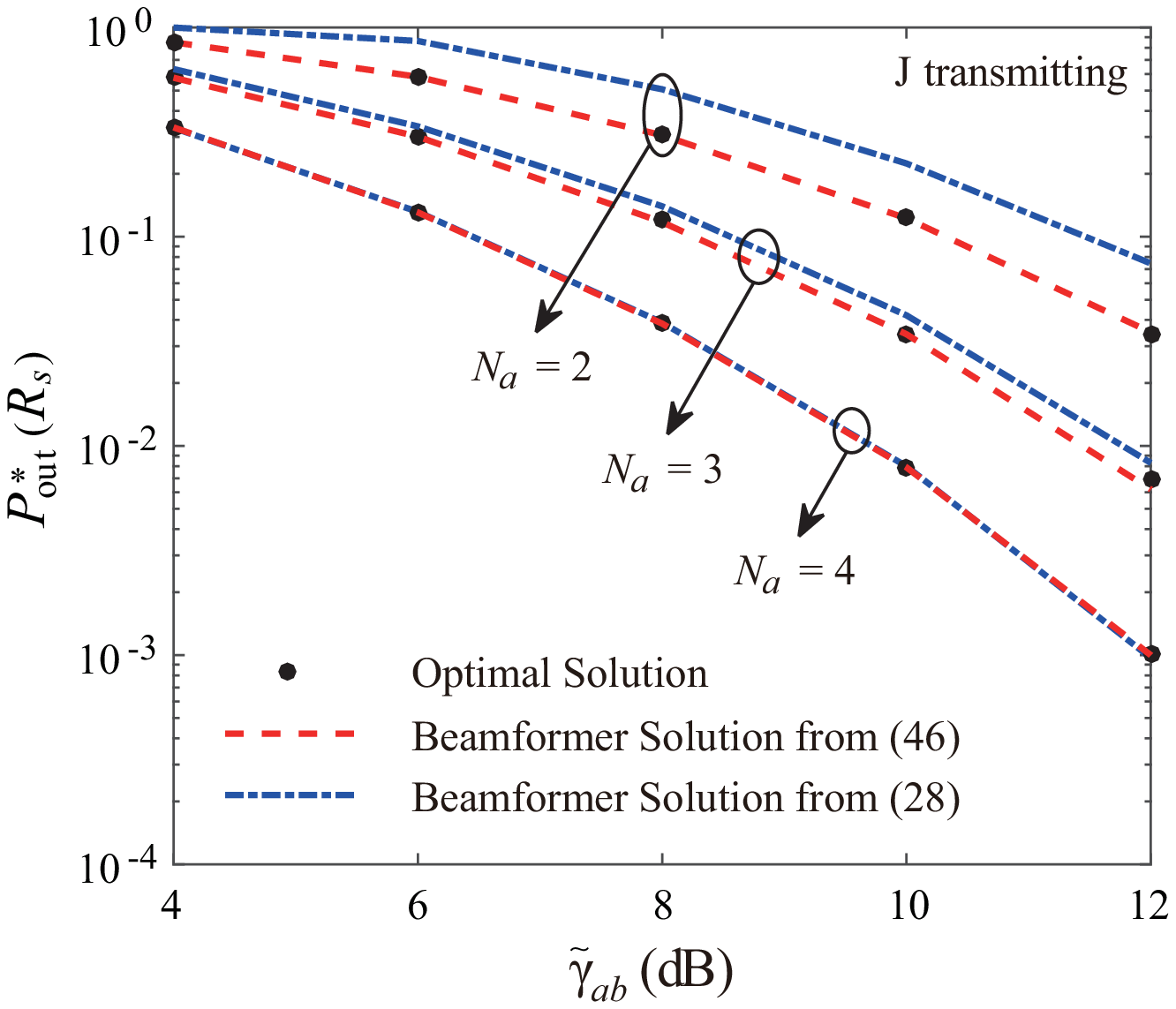}}
\caption{$P_{\text{out}}^{\ast}\left(R_s\right)$ versus $\tilde{\gamma}_{ab}$ for different values of $N_a$ with $N_e=2$, $N_j=4$, $K_{ab}=10$ dB, $K_{ae}=5$ dB, $K_{je} = 0$, $\tilde{\gamma}_{ae} = \tilde{\gamma}_{je}= 10$ dB, $\theta_{ab}=\pi/3$, $\theta_{ae} = \pi/4$, and $R_s=1$ bits/s/Hz. J is transmitting here.}\label{fig_side_e}
\end{center}
\end{figure}
We then examine the effectiveness of the scheme when J is transmitting in the following Figs. \ref{fig_side_c}--\ref{fig_side_e}. To provide focus, we consider the special case of $K_{je}=0$.
In Fig. \ref{fig_side_c}, we plot $P_{\text{out}}\left(R_s\right)$ versus $\tau$ for different values of $N_a$. We first see that the analytical curves, generated from {\em Proposition \ref{p2}} and {\em Theorem \ref{t2}}, effectively match the simulation points for $N_a>2$. We clarify that the gap between the analytical curve and simulation points when $N_a=2$ is due to the fact that we adopt a gamma approximation method to characterize the distribution of the received SINR at Eve. We note that, the small gap between the analytical curve and the simulations when $N_a=2$ is a constant, revealing that we can still use {\em Proposition \ref{p2}} and {\em Theorem \ref{t2}} to determine the optimal $\tau^{\ast}_j$ that minimizes $P_{\text{out}}\left(R_s\right)$ even when $N_a=2$. We then see that there  exists a unique $\tau^{\ast}_j$ that minimizes $P_{\text{out}}\left(R_s\right)$ for each $N_a$. We also see that the optimal $\tau^{\ast}_j$ that minimizes $P_{\text{out}}\left(R_s\right)$ decreases for each $N_a$, compared to the optimal $\tau^{\ast}$ that minimizes $P_{\text{out}}\left(R_s\right)$ in Fig. \ref{fig_side_a}. This is due to the fact that the jamming signals degrade the quality of the received signals at Eve.



In Fig. \ref{fig_side_e}, we plot $P^{\ast}_{\text{out}}\left(R_s\right)$ versus $\tilde{\gamma}_{ab}$ for different values of $N_a$. In this figure, we compare the secrecy performance of the beamformer solution obtained from {\em Proposition \ref{p1}} and {\em Theorem \ref{t1}} and the beamformer solution obtained from {\em Proposition \ref{p2}} and {\em Theorem \ref{t2}} to the secrecy performance of the optimal beamformer solution. The beamformer solution obtained from {\em Proposition \ref{p1}} and {\em Theorem \ref{t1}} and the beamformer solution obtained from {\em Proposition \ref{p2}} and {\em Theorem \ref{t2}} are represented by blue-dashed dotted lines and red-dashed lines, respectively. The optimal solutions, represented by `$\bullet$' symbols, are obtained from minimizing $P_{\text{out}}\left(R_s\right)$ via an exhaustive search for different values of $N_a$.  Similar as in Fig. \ref{fig_side_b}, we first observe that the minimum secrecy outage probability $P_{\text{out}}^{\ast}\left(R_s\right)$ achieved by the beamformer solution from {\em Proposition \ref{p2}} and {\em Theorem \ref{t2}} is nearly the same as the optimal beamformer solution found through exhaustive search. We then observe that $P^{\ast}_{\text{out}}\left(R_s\right)$ decreases significantly as $N_a$ decreases and $\tilde{\gamma}_{ab}$ increases. Moreover, we observe that the gap between the minimum $P_{\text{out}}\left(R_s\right)$ achieved by the beamformer solution from {\em Proposition \ref{p1}} and {\em Theorem \ref{t1}} and the minimum $P_{\text{out}}\left(R_s\right)$ achieved by the beamformer solution from {\em Proposition \ref{p2} and Theorem \ref{t2}} reduces as $N_a$ increases, revealing that, in Rician wiretap channels with the jammer, the secrecy performance of the beamformer solution obtained from {\em Proposition \ref{p1}} and {\em Theorem \ref{t1}} (i.e., from \textbf{Algorithm \ref{a1}}) is almost the same as the secrecy performance of the beamformer solution obtained from {\em Proposition \ref{p2}} and {\em Theorem \ref{t2}} when $N_a$ is larger than 2.
%

\begin{figure}[!t]
    \centering
    \subfigure[$K_{ae} = 0$.]
    {
        \includegraphics[height=1in,width=1in]{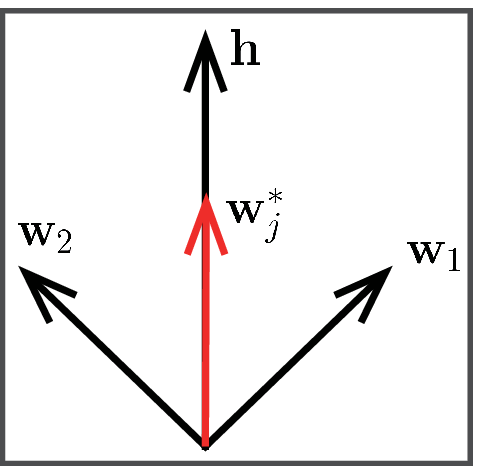}
        \label{fig:first_sub}
    }
    \subfigure[$K_{ae} = \infty$.]
    {
        \includegraphics[height=1in,width=1in]{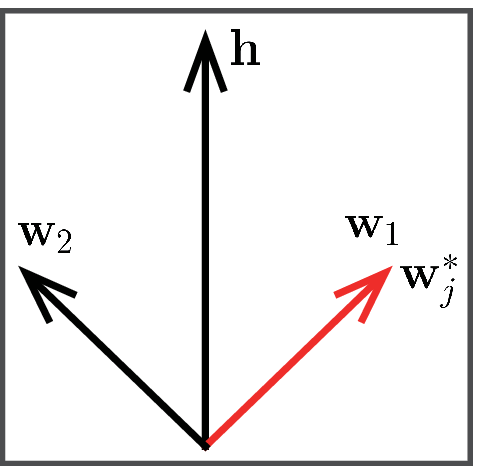}
        \label{fig:second_sub}
    }
    \subfigure[$K_{ae} = 5$ dB. ]
    {
        \includegraphics[height=1in,width=1in]{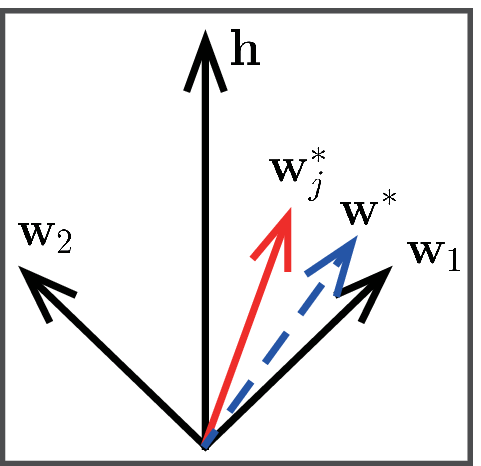}
        \label{fig:third_sub}
    }
    \caption{Illustration of the optimal beamformer solutions for Rician wiretap channels with $N_a = 2$, $N_e=2$, $K_{ab}=10$ dB, $K_{je} = 0$, $\tilde{\gamma}_{ab}=\tilde{\gamma}_{ae} = \tilde{\gamma}_{je}= 10$ dB. $\mathbf{w}^{\ast}$ and $\mathbf{w}_j^{\ast}$ denote the optimal beamformer solution when J is not transmitting and the optimal beamformer solution when J is transmitting, repectively. $\mathbf{w}^{\ast}$ and $\mathbf{w}_j^{\ast}$ are represented by blue dashed line and red solid line, respectively.}
    \label{fig:sample_subfigures}
\end{figure}

In Fig. \ref{fig:sample_subfigures}, we provide a schematic view of the optimal beamformer solutions for different values of $K_{ae}$. For illustration purpose, we denote $\mathbf{w}^{\ast}$ as the optimal beamformer solution when J is not transmitting. We also denote $\mathbf{w}_j^{\ast}$ as the optimal beamformer solution when J is transmitting. Fig. \ref{fig:first_sub} shows the optimal beamformer solutions when $K_{ae}=0$ (i.e, the Alice-Eve channel is in a pure Rayleigh fading environment). We see that $\mathbf{w}^{\ast}$ and $\mathbf{w}_j^{\ast}$ overlap with each other. We also see that $\mathbf{w}^{\ast}$ and $\mathbf{w}_j^{\ast}$ are in the same direction as the main channel $\mathbf{h}_{ab}$, indicating that the optimal beamformer solutions when $K_{ae}=0$ are the maximal-ratio transmission such that the capacity of the main channel is maximized. Fig. \ref{fig:second_sub} shows the optimal beamformer solutions when $K_{ae}=\infty$ (i.e., the Alice-Eve channel is in a pure LOS environment). We see that $\mathbf{w}^{\ast}$ and $\mathbf{w}_j^{\ast}$ overlap with $\mathbf{w}_1$, revealing that the optimal beamformer solutions are orthogonal to the LOS component in the Alice-Eve channel $\mathbf{G}_{ae}^{o}$. Moreover, we examine more general scenarios where $K_{ae}$ is between the above two extremes. As a specific example, in Fig. \ref{fig:third_sub} we show the optimal beamformer solutions of our scheme for $K_{ae} = 5$ dB. We first see that $\mathbf{w}^{\ast}$ and $\mathbf{w}_j^{\ast}$ are in different directions, which validates our analysis in {\em Theorem \ref{t2}}.
We then see that $\mathbf{w}_j^{\ast}$ is closer to the main channel $\mathbf{h}_{ab}$, compared to $\mathbf{w}^{\ast}$. This is due to the fact that the jamming signals degrade the quality of the received signals at Eve.

%
We now examine the impact of the uncertainty in Eve's location. To this end, we adopt the time difference of arrival (TDOA) scheme  discussed in \cite{chenxi4} as the location estimation scheme. We then introduce the covariance matrix $\mathbf{V}_{pos} = \mathbf{J}^{-1}$, where J denotes the Fisher matrix for TDOA scheme (see \cite{chenxi4} for details). We further express $\mathbf{V}_{pos}$ as
\begin{align}
\label{v_pos_2} \mathbf{V}_{pos} = \left[
\begin{array}{*{20}{c}}
  \sigma_{x}^2&\sigma_{xy}\\
  \sigma_{yx}&\sigma_{y}^2
\end{array}\right],
\end{align}
where the values of $\sigma_{x}$, $\sigma_{y}$, $\sigma_{xy}$, and $\sigma_{yx}$ can be obtained straightforwardly from the inverse of $\mathbf{J}$. We denote Eve's true location as ${\zeta}_0 = \left[{x}_0,{y}_0\right]$, Eve's estimated location as ${\zeta}_e = \left[{x}_e,{y}_e\right]$, and the correlation coefficient as
$
\rho = \sigma_{xy}/\left({\sigma_x\sigma_y}\right).
$
As such, the distribution of Eve's estimated location can be expressed as
\begin{align}
\label{p_xy} P({\zeta}_e)=& \dfrac{1}{2\pi\sqrt{1-\rho^2}\sigma_x\sigma_y}\exp\left\{-\dfrac{1}{2\left(1-\rho^2\right)}\left(\dfrac{\left({x}_e-{x}_0\right)^2}{\sigma_x^2}\right.\right.\notag\\
&\left.\left.+\dfrac{\left({y}_e-{y}_0\right)^2}{\sigma_y^2}
-2\dfrac{\rho\left({x}_e-{x}_0\right)\left({y}_e-{y}_0\right)}{\sigma_x\sigma_y}\right)\right\}.
\end{align}
In order to characterize the secrecy performance of the system, we adopt an ``average'' measure of $P_{\text{out}}\left(R_s\right)$, which is given by \cite[Eq.(44)]{chenxi4}
\begin{align}
\label{average_outage} \overline{P}_{\text{out}}\left(R_s\right) = \int_{-\infty}^{\infty}\int_{-\infty}^{\infty}{P}_{\text{out}}\left(R_s\right)P\left(\xi_e\right)dx_edy_e.
\end{align}

\begin{figure}[t!]
\begin{center}{\includegraphics[height=2.8in,width = 3.0in]{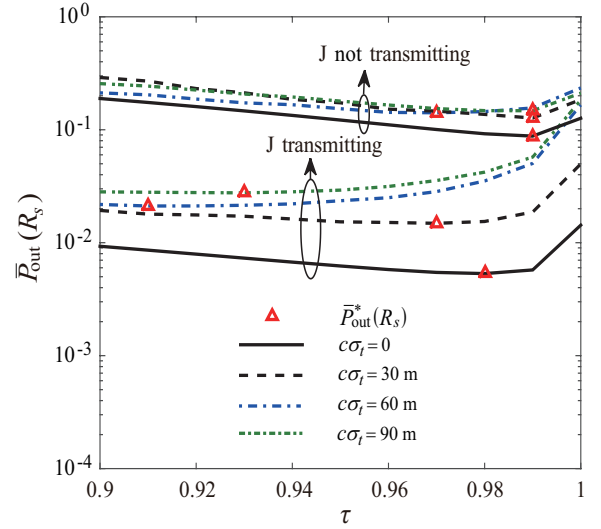}}
\caption{$\overline{P}_{\text{out}}\left(R_s\right)$ versus $\tau$ for different values of $c\sigma_t$ with $N_a=4$, $N_e=2$, $N_j = 4$, $K_{ab}=10$ dB, $K_{ae}=5$ dB, $\tilde{\gamma}_{ab}=\tilde{\gamma}_{ae}= 10$ dB, $\theta_{ab}=\pi/3$, $\theta_{ae} = \pi/4$, and $R_s=1$ bits/s/Hz.}\label{fig_side_g}
\end{center}
\end{figure}

In Fig. \ref{fig_side_g}, we plot $\overline{P}_{\text{out}}\left(R_s\right)$ versus $\tau$ for different levels of Eve's location uncertainty for both the case where J is not transmitting and the case where J is transmitting. In this figure, we adopt the TDOA scheme  discussed in \cite{chenxi4} as the location estimation scheme. In the TDOA scheme, the level of Eve's location uncertainty is represented by $c\sigma_t$, where $c$ is the speed of the light, and $\sigma_t$ is the standard deviation of the timings. The larger $c\sigma_t$ is, the less accurate Eve's location is. In Fig. \ref{fig_side_g},  we consider that Alice, Bob, and J are located at $[0\text{m},~0\text{m}]$, $[1225\text{m},~707\text{m}]$, and $[2000\text{m},~-3464\text{m}]$, respectively. We also consider that the true location of Eve is $[1000\text{m},~-1000\text{m}]$ (note the coordinates are chosen according to set angles). We clarify that we choose this parameter setting to mimic scenarios where the distance between nodes is relatively large. We see that, for both cases, there exists a unique $\tau^{\ast}$ that minimizes $\overline{P}_{\text{out}}\left(R_s\right)$ for each $c\sigma_t$. We also see that the minimum $\overline{P}_{\text{out}}\left(R_s\right)$ increases as $c\sigma_t$ increases, which demonstrates that the secrecy performance of our scheme decreases, as the level of uncertainty in Eve's location increases. Although not completely shown here, we note that our  results approach the appropriate solutions as the location uncertainty approaches both zero and infinity (i.e., location unknown), and show the expected trends between these two extremes. Moreover, compared to the case where J is not transmitting, we can observe that, for a specific $c\sigma_t>0$, the minimum $\overline{P}_{\text{out}}\left(R_s\right)$ of the case where J is transmitting is decreased (as expected). We note that similar trends and outcomes to those shown in Fig. \ref{fig_side_g} were found for a wide range of antenna configurations, transceiver locations, and Eve locations.

\section{Conclusion}
In this work we have proposed a new LBB solution for Rician wiretap channels, in which a source communicates with a legitimate receiver in the presence of an eavesdropper.
In our scheme, we assumed that the CSI from the legitimate receiver is known at the source, while the only available information on the eavesdropper at the source is her location. With no jammer present, we showed how the beamforming vector that minimizes the secrecy outage probability of the system can be obtained in real-time. We also examined the optimal beamformer solution  in the presence of a multi-antenna jammer, showing how our real-time no-jammer solution still provides close-to-optimal performance in most practical scenarios.
 The work reported here illustrates how in  a range of realistic wiretap channels, in which the only information known on an eavesdropper is her location, a real-time solution to the optimal beamformer can be determined and deployed.

\begin{appendices}
\section{Proof of Theorem \ref{t2}}\label{App_t2}
In order to derive the secrecy outage probability for Rician fading wiretap channels with a jammer for the special case of $K_{je}=0$, we first need to derive the PDF of $\gamma_{e}$. We note that the closed-form expression for the PDF of $\gamma_{e}$ is mathematically intractable due to the fact that $\gamma_{e}$ is a random variable containing both the non-central complex normal vector $\mathbf{G}_{ae}\mathbf{w}$ and the random matrix $\mathbf{R}$. To address this problem, we consider the use of the gamma approximation to characterize the PDF of $\gamma_{e}$. Such an approximation has been shown to be effective in accurately describing the distribution of the received SINR of Rician fading channels with Rayleigh-distributed co-channel interference \cite{raymond}. As such, we express the gamma approximations of the PDF $\gamma_{e}$ as
\begin{align}
\label{PDF_snr_e_j} f_{\gamma_{e}}\left(x\right) =
\frac{x^{\alpha-1}\exp\left(-\frac{x}{\beta}\right)}{\Gamma\left(\alpha\right)\beta^{\alpha}},
\end{align}
where $\alpha$ denotes the scale parameter of the gamma distribution, and $\beta$ denotes the shape parameter of the gamma distribution. We have $\alpha\beta$ and $\alpha\beta^2$ represent the mean and the variance of $\gamma_{e}$, respectively.
We then express the CDF of $\gamma_{e}$ as
\begin{align}\label{CDF_snr_e_j}
F_{\gamma_{e}}\left(\gamma\right) = \frac{\gamma\left(\alpha,\frac{x}{\beta}\right)}{\Gamma\left(\alpha\right)},
\end{align}

We express the $l$th moment of $\gamma_{e}$ as \cite{mckay}
\begin{align}
\label{xi}  \xi_l = \tilde{\gamma}_{ae}^l\varphi_l\vartheta_l,
\end{align}
where $\varphi_l$ and $\vartheta_l$ are as shown in \eqref{varphi_l} and \eqref{vartheta_l}, respectively.
Based on \eqref{xi}, we obtain the mean of $\gamma_{e}$ as
\begin{align}
\label{mean} \mathbb{E}\left[\gamma_{e}\right] = \tilde{\gamma}_{ae}\varphi_1\vartheta_1.
\end{align}
We then obtain the variance of $\gamma_{e}$ as
\begin{align}\label{variance}
\text{Var}\left(\gamma_{e}\right) = \tilde{\gamma}_{ae}^2\left(\varphi_2\vartheta_2-\varphi_1^2\vartheta_1^2\right).
\end{align}
From \eqref{mean} and \eqref{variance}, we obtain $\alpha$ and $\beta$ as
\begin{align}
\label{alpha} \alpha = \frac{\varphi_1^2\vartheta_1^2}{\varphi_2\vartheta_2-\varphi_1^2\vartheta_1^2},
\end{align}
and
\begin{align}
\label{beta} \beta = \tilde{\gamma}_{ae}\frac{\left(\varphi_2\vartheta_2-\varphi_1^2\vartheta_1^2\right)}{\varphi_1\vartheta_1},
\end{align}
respectively.

We then re-express $P_{\text{out}}\left(R_s\right)$  in \eqref{secrecy_outage_probability} as
\begin{align}
\label{secrecy_outage_probability_5} P_{\text{out}}\left(R_s\right) &= \mbox{Pr}\left(C_b-C_{e}<R_s|\gamma_b\right)\notag\\
&=\mbox{Pr}\left(C_{e}>C_b-R_s|\gamma_b\right)\notag\\
&=\mbox{Pr}\left(\gamma_{e}>2^{-R_s}\left(1+\gamma_b\right)-1\right)\notag\\
&=1-\int_{0}^{2^{-R_s}\left(1+\gamma_b\right)-1}f_{\gamma_{e}}\left(\gamma\right)d\gamma\notag\\
&=1-F_{\gamma_{e}}\left(2^{-R_s}\left(1+\gamma_b\right)-1\right).
\end{align}
Substituting \eqref{PDF_snr_e_j}, \eqref{CDF_snr_e_j}, \eqref{alpha}, and \eqref{beta} into \eqref{secrecy_outage_probability_5}, we obtain the desired result in \eqref{t2_result}. The proof is completed.
\end{appendices}

\end{document}